\documentclass[letterpaper]{aipproc}

\layoutstyle{8x11double}

\begin{document}
\author{M. Barisonzi}{address={University of Twente, Postbus 217,  
7500 AE Enschede, The Netherlands\\NIKHEF, Kruislaan 409, 1098 SJ Amsterdam, The Netherlands}}
\title{Top Physics at ATLAS}

\begin{abstract}
The Large Hadron Collider {\bf LHC} is a top quark factory: due to its high design luminosity, LHC will produce about 200 millions of top quarks per year of operation. The large amount of data will allow to study with great precision the properties of the top quark, most notably cross-section, mass and spin. The Top Physics Working Group has been set up at the ATLAS experiment, to evaluate the precision reach of physics measurements in the top sector, and to study the systematic effects of the ATLAS detector on such measurements. This reports give an overview of the main activities of the ATLAS Top Physics Working Group in 2004.
\end{abstract}

\maketitle

%\tableofcontents

\section{Overview of the Top Quark}
The top quark is the fermion with the largest mass in the Standard Model. In the Higgs mechanisms, this means that the top quark has a large coupling strength with the Higgs boson.
This property can be exploited to constrain the expected mass of the Higgs boson. A NLO calculation of the mass of the $W$ boson contains corrections term dependent on the top mass and the Higgs mass. By accurately measuring $M_W$ and $m_t$, it is possible to obtain an estimate for the $M_H$. Recent Tevatron measurements for the top mass of 178.0\textpm4.3~GeV result in a Higgs mass prediction of 126$^{+73}_{-48}$~GeV.\par
The achievement of a high top mass resolution is instrumental for a precise prediction of $M_H$.

\subsection{Production processes}
Top quarks are produced at LHC by two types of processes: QCD and Electroweak production. In QCD production, top quarks are created in $t\bar{t}$ pairs via the processes $q\bar{q}\to t\bar{t}$ and $gg\to t\bar{t}$. The combined NLO cross-section for these two processes is $\sim$825~pb; 90\% of the contribution comes from the gluon-gluon fusion process, the remaining 10\% comes from quark-antiquark annihilation.\par
Electroweak production is composed of three channels: the s-channel, the t-channel (also referred as $W$-gluon fusion) and associated $Wt$ production. The three processes have cross sections of 11~pb, 60~pb, and 247~pb respectively.% the cross-section of the $W$-gluon fusion channel is of the same order of magnitude as the QCD pair production.
The cross-section for EW processes is directly proportional to the CKM matrix element $V_{tb}$: a deviation from the SM prediction may point to the existence of a fourth quark generation.\par

\subsection{Experimental signatures}
In the Standard Model, the top quark decays predominantly into a $W$ and a $b$-quark with a branching ratio of 0.998. Because of fermion universality in electroweak interactions, the $W$ boson decays 1/3 of the time into a lepton/neutrino pair and 2/3 of the time into a $q\bar{q}$ pair. Since in $t\bar{t}$ events two real $W$ bosons are present, the signatures of the events are classified according to the decays of the two $W$ bosons: 

\begin{description}
\item[all jets channel] In this channel both $W$s decay into a quark/anti-quark pair. The event has at least six high-$p_T$ jets, two of which have to be b-tagged. Despite having the highest branching ratio (44\%), this decay channel suffers heavily from QCD background and ambiguities in the assignment of jets to the originating $W$s.
\item[lepton+jets channel] In this decay channel one $W$ decays into a lepton--neutrino pair, the other $W$ into a quark/anti-quark pair. Thus, the top and anti-top quarks are called ``leptonic'' or ``hadronic'' accordingly with the decay mode of their daughter $W$. The event signature is characterised by one isolated lepton one b-jet and missing $E_T$ in the leptonic branch, two light jets and one b-jet in the hadronic branch. The branching ratio for this channel is about 30\%. 
\item[di-lepton channel] In this decay channel, both $W$s decay into a lepton--neutrino pair. For practical purposes, only $e,\mu$ are considered, since $\tau$ decays are difficult to distinguish from the QCD background. The events have two high-$p_T$ leptons, two jets (at least one of which is b-tagged) and missing energy due to the neutrinos. This signature is quite clear, being affected mainly by electroweak  background. The only drawback of this decay channel is its low branching ratio (5\%).
\end{description}

The lifetime of the top quark is much shorter than the time constant for spin-flip via gluon emission. Thus, decay products retain the spin information of the original top quark. By measuring the spin of the decay products, it is possible to verify the $V-A$ coupling of the $Wtb$ vertex or discover a non-SM right-handed coupling.

%% Several experimental signatures from the decay of the two top quarks:
%% The W decays either into a lepton and a neutrino (BR~33\%) or into a qq' pair (BR~66\%)
%% The top quark is then called ``leptonic'' or ``hadronic'' accordingly with the decay mode of its daughter W.
%% Leptonic W  reconstructed by combining lepton and missing energy, which approximates the neutrino.
%% The pz of the neutrino can be evaluated using momentum conservation:                                                    
%% The b-jet is short-lived and produces a secondary vertex (b-tagging) 
%% Hadronic W reconstructed by combining two hadronic jets (usually highly energetic)
%% Stumbling blocks: b-tag efficiency, missing ET, combinatorics

\section{Mass measurements}
The lepton+jets $t\bar{t}$ channel is the most promising for top mass measurement; however the channel is affected by combinatorial background. The assignment of the two b-jets to the hadronic and leptonic branches is not unique, and can spoil the resolution of the top mass measurement.\par
A study was performed to evaluate the best strategy for the assignment of b-jets. The study was performed on a MC sample of 200,000 $t\bar{t}$ events with full detector simulation. The selection cuts required at least 4 Jets of $p_T$>20~GeV and $|\eta|$<2.5, $E_T^{miss}$>20~GeV and only one isolated lepton of $p_T$>20~GeV. In addition, two b-tagged jets were required with varying $p_T$ thresholds.\par
The assignment algorithm proceeded as follows: the hadronic branch is assigned to the b-jet with the minimum distance $\Delta R$ from the hadronic $W$, while the leptonic branch is assigned to the b-jet with the minimum azimuthal angle $\phi$ from the leptonic $W$. If the algorithm produces conflicting assignment, then the algorithm chooses the basis of another variable: the maximum $\Delta R$ between the b-jet and the lepton, the combination that maximizes $p_T(Top_{had})+p_T(Top_{lep})$ or the combination that minimizes $\phi(b_i,W_{lep})+\phi(b_j,W_{had}$ or one of the two angles $\phi(b_i,W_{lep})$, $\phi(b_j,W_{had})$.\par
\begin{figure}[htb]
\resizebox{0.45\textwidth}{!}
  {\includegraphics{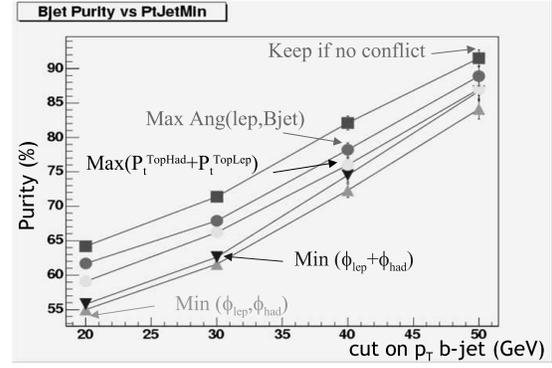}}
\caption{Purity in the $W-b$ assignment for various assignment methods over varying $p_T$ threshold for the b-jets.}
\label{fig:etienvre}
\end{figure}

For each of the variables used in the assignment algorithm, the purity of the assignment was computed. The purity varies according to the cut on the minimum $p_T$ of the b-jets: the higher the threshold, the better the purity achieved (Figure~\ref{fig:etienvre}). On the other hand, a high threshold biases the mass of the reconstructed top quark towards higher value. The best compromise is to choose a $p_T$ threshold for b-jets of 40~GeV, which results in an assignment purity of 70-80\%.

\section{Spin correlations}
Top/anti-top quark pairs produced in QCD processes are not polarized; however, the top and anti-top spins are correlated. For QCD processes close to production threshold, the $t\bar{t}$ system is prodeuced in a $^3S_1$ state for $q\bar{q}$ annihilation, or in a $^1S_0$ state for gluon-gluon fusion. Hence, in the first case the top and the anti-top have parallel spins, while in the second case the spins are antiparallel. Since at LHC the gluon-fusion process has a much larger cross-section than $q\bar{q}$ annihilation, NLO calculations predict an excess of top pairs with opposite spins. The spin of the quarks can be evaluated in the helicity basis, which corresponds to the top (anti-top) direction of flight in the $t\bar{t}$ system. In this basis, the asymmetry parameter --- that expresses the excess of same-helicity pairs --- is given by:

\begin{eqnarray*}
A & = &\frac{\sigma(t_L\bar{t}_L)+\sigma(t_R\bar{t}_R)-\sigma(t_L\bar{t}_R)-\sigma(t_R\bar{t}_L)}{\sigma(t_L\bar{t}_L)+\sigma(t_R\bar{t}_R)+\sigma(t_L\bar{t}_R)+\sigma(t_R\bar{t}_L)}\\
  & = & 0.326_{-0.002}^{+0.003} \left( \mu_{r,f} \right) _{-0.001}^{+0.013} \left( PDF \right)
\end{eqnarray*}

For $t\bar{t}$ pairs produced with total invariant mass much larger than production threshold, the asymmetry is diluted because of the presence of higher spin $t\bar{t}$ pairs. Hence it is useful to introduce a cut in the total invariant mass $m_{t\bar{t}}<550$~GeV to maximize the power of the experimental analysis. A significant deviation of the measured asymmetry parameter from the theoretical value may point to non-SM physics, such as right-handed weak interactions.\par
The asymmetry can be evaluated by studing the angular distributions of the top and anti-top decay products (spin analyzers):

\begin{eqnarray*}
 {1 \over N}\frac{d^2N}{d\cos\theta_1 d\cos\theta_2} & = & {1 \over 4} \left( 1 - C\cos\theta_1 \cos\theta_2 \right) \\
 {1 \over N}\frac{dN}{d\cos\phi} & = & {1 \over 2}\left(1 - D\cos\phi \right)
\end{eqnarray*}

where $\theta_1$ ($\theta_2$) indicates the the angle between the decay product of $t$ ($\bar{t}$) in the $t$ ($\bar{t}$) rest frame and the  $t$ ($\bar{t}$) direction in the $t\bar{t}$ frame, while $\phi$ indicates the angle between the decay products of $t$ and $\bar{t}$ in the respective rest frames \cite{spincorr}.

%Analysing power:
%$${dN \over d\cos\theta_i} \sim 1+\alpha_i\cos\theta_i$$

The parameters $C$ and $D$ are the spin correlation variables, which can be evaluated using the following unbiased estimators:

\begin{eqnarray*}
C & = & -9\langle\cos\theta_1\cos\theta_2\rangle \\
D & = & -3\langle\cos\phi\rangle
\end{eqnarray*}

The spin correlation parameters have been measured on MonteCarlo data in the dileptonic and the lepton+jets decay channels. For the dileptonic channel, the natural choice for spin analyzers is the two leptons from the decay of the $W$s, since leptons are 100\% polarized with respect to the top spin. In the lepton+jets channel, the best choice for spin analyzers would be a lepton and a $d$-type quark, since $d$ have 100\% polarization as well. However, since it is experimentally impossible to distinguish between light quark flavours, the second spin analyzer is the least energetic jet in the $t$ ($\bar{t}$) rest frame, which is polarised at 51\%.\par
The $t\bar{t}$ sample was created using the event generator TopReX~4.05 \cite{toprex} in combination with the hadronisation software Pythia~6.2 \cite{pythia} and the fast detector simulation ATLFAST \cite{atlfast}. The background sample included $t\bar{t}$ events decaying to all-hadrons and $\tau$+jets, $W$+jets, $Z$+jets, $Wb\bar{b}$, $ZZ$, $WW$, $WZ$ and single top production. The total generated sample corresponds to 10~fb$^{-1}$, or one year at LHC at low luminosity; the sample was generated with varying production parameters to study the systematic effects of:

\begin{itemize}
\item Q-scales;
\item Parton Distribution Functions;
\item Initial and Final State Radiation;
\item b-fragmentation and hadronisation scheme;
\item b-tagging and b-jet calibration;
\item generated top mass.
\end{itemize}

The results of the spin correlation measurements in the two channels are summarised in Table~\ref{tab:spincorr}; the spin correlation parameter $D$ resulted more suitable for measurements, since it is less sensitive to systematic effects.

\begin{table}
\begin{tabular}{lcc}
\hline
Parameter & Di-lepton & Lepton+jets \\
\hline
{\bf C} (theory) & 0.42 & 0.21\\
{\bf C} (MC)     & 0.40\textpm{\em 0.02}\textpm0.02 & 0.21\textpm{\em 0.01}\textpm0.04 \\
Precision        & 8\% & 20\% \\
\hline
{\bf D} (theory) & -0.29 & -0.15 \\
{\bf D} (MC)     & -0.29\textpm{\em0.01}\textpm0.01 & -0.15\textpm{\em 0.006} \textpm0.02\\
Precision        &  5\% & 13\% \\
\hline
\end{tabular}
\caption{\label{tab:spincorr} Comparison between the theoretical predictions and the MC measurements for spin correlation parameters $C$ and $D$ in the di-leptonic and the lepton+jets sample. The statistical error is written in italics, while the systematic error is written in plain typeface.}
\end{table}

\section{Detector commissioning studies}
The top pair production process is valuable for the in-situ calibration of ATLAS in the detector commissioning stage: the large cross section and the large $S/B$ ratio for the lepton+jets $t\bar{t}$ channel allow to produce high purity samples with large statistics in a short time period. The experimental signature for top events --- high energy leptons and jets, $b$-jets, missing $E_T$ --- involves most parts of the ATLAS detector, so top samples can play an important role in the calibration of the detector in the initial phases.\par

\subsection{Top analysis without b-tagging} 
The experimental signature for top events include one or more b-jets, arising from the top decay. Thus, the b-tagging performance of ATLAS has an important role for top analyses. However, efficient b-tagging needs precise alignment of the trackers of the Inner Detector, which will be reached only after few months of data taking. Will it be possible to perform top physics in the commissioning stage? A study has explored the possibility of reconstructing top events in $t\bar{t}$ production by assuming the absence of b-tagging.\par
The kinematic cuts applied to select the top-enriched sample require exeactly one isolated lepton with $p_T$>20~GeV and exactly four jets --- reconstructed with a cone algorithm of size $\Delta R$=0.4 --- with $p_T$>40~GeV. The analysis reconstructs exclusively the hadronic branch of the $t\bar{t}$ event: the combination of 3 out of 4 jets with the highest vector $p_T$ sum is assumed to belong to the same top (antitop) branch. The invariant mass of the 3-jets combination is an estimate of the top mass.\par
\begin{figure}[htb]
\resizebox{0.45\textwidth}{!}
  {\includegraphics{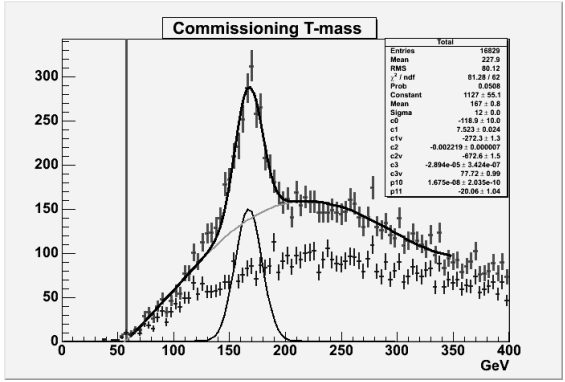}}
\caption{Invariant mass distribution of the 3-jets combinantion with the highest vector $p_T$ sum. The full curve fits the signal+background ($W$+jets and combinatorial), while the lower distribution shows the contribution of $W$+jets only.}
\label{fig:stan}
\end{figure}

The analysis was performed on a generated data sample which included the $t\bar{t}$ signal plus a $W$+4jets background (leptonic decay of $W$ with 4 extra light jets). The invariant mass distribution (Figure~\ref{fig:stan}) is fitted with a gaussian curve --- describing the top peak mass --- plus a polynomial curve --- which accounts for the $W$+jets signal plus the combinatorial background. The fitted gaussian curve peaks at 167.0~GeV, with a RMS of 12~GeV.\par
The generated sample corresponds to 150~pb$^{-1}$, i.e. a few days of data taking at initial luminosity. This study shows that top reconstruction can be performed in absence of b-tagging.

\subsection{$W$ calibration from top decays}
In order to obtain a precise measurement of the mass of the top quark, it is necessary to know the absolute energy scale of hadronic jets. Miscalibrations can arise from detector effects (dead channels, imprecise cell weighing), physics effects (final state radiation, pile-up), cone algorithms effects (out-of-cone energy, jet overlap). The best method to achieve a reliable calibration is to use the physics data to reconstruct particles of known properties. One of this techniques proposes to reconstruct the hadronic $W$ from the lepton+jets $t\bar{t}$ sample. %The distinctive signature of this sample guarantees a high sample purity 
The kinematic cuts used to select the sample are the same as the ``commissioning top'' cuts described in the previous section. In addition, a quality cut is applied: the 4-vectors of the fourth jet, the lepton and the $E_T^{miss}$ are summed, and the invariant mass is required to be in the range 140~GeV<$M_{j_4\ell\nu}$<200~GeV. With this set of cuts, the purity for $W$ in the selected sample is $\sim$85\%.\par
\begin{figure}[htb]
\resizebox{0.45\textwidth}{!}
  {\includegraphics{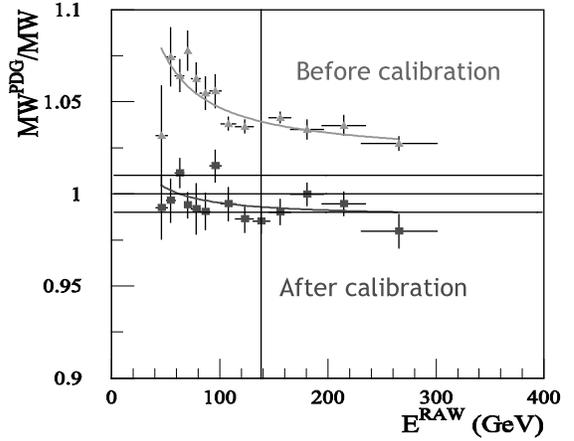}}
\caption{Ratio between the nominal and reconstructed $W$ mass, as a function of the uncalibrated $W$ energy, before and after the jet scale calibration.}
\label{fig:pallin}
\end{figure}

  
Since the mass of the $W$ is known to a precision of a few MeV, $M_W$ can be used to obtain calibration factors for the energies of the jets originating from the decay of the $W$. For the jets $j_1,j_2$ from the $W$ decay, momentum conservation implies:
$$M_W^2=2E_{j_1}E_{j_2}\left( 1 - \cos\theta_{j_1j_2} \right)$$
where $E_{j_i}$ indicates the energy of jet $i$ and $\theta_{j_1j_2}$ indicates the angle between the two jets. Thus, the mass measurement is influenced both by the resolution on the jet energy scale and the angular measurement. To disentangle the energy and the angular contributions, the invariant mass of the two-jets system, the $W$ nominal mass and the kinematic properties of the jets are used as input in a constrained $\chi^2$ fit:
$$\chi^2 = \left({ m_{jj}  - M_W} \over {\sigma_{M_W}}  \right)^2  + \sum_{i,X} \left[ {{X_i  - \alpha_E^i X_i }  \over {\sigma_X }}  \right]^2$$
where $X=E,\eta,\phi$. The result of the fitting procedure is the correction factor $\alpha_E$, which is assumed to depend on the jet energy $E$. By plotting $\alpha_E$ versus the jet energy, a correction function $\alpha_E=f(E)$ is obtained. Applying the correction factor allows to reconstruct the $W$ mass with a 1\% precision (Figure~\ref{fig:pallin}). To obtain such a precision level, only 10000 $t\bar{t}$ events are necessary --- that means that the calibration will be achieved within one or two months of data taking. The calibration factors have been applied to a $Z$+jets sample, allowing the same level of precision and showing that the results of the calibration can be used for other physics signatures involving light jets.

%% Top quark detectable already in the first weeks of data taking
%% b-tagging not strictly necessary in the commissioning phase
%% Hadronic decays of W from ttbar events useful for jet energy calibration
%% W mass resolution under 3\% after max. 3 weeks

%% \section{Conclusions}
%%  Accurate top mass measurements are necessary before we can start   thinking about the Higgs

%%  ATLAS Top Physics WG very active in the study of systematics which can affect the mass measurement
 
%%  ttbar events are an important tool to understand our detector:
%%  ``independence''     from b-tagging ?? visible from Day 1
%%  W from top decay useful for jet scale calibration
%%  But! Most studies are still performed at LO, with Fast Simulation
%%  Sizeable set of MC samples with Full Simulation available from May
%%  MC@NLO available for ttbar right now, available for single top in Summer 2005 

\end{document}